\documentclass[twocolumn,showpacs,preprintnumbers,amsmath,amssymb,nofootinbib]{revtex4-1}
\usepackage{graphicx}% Include figure files
\usepackage{dcolumn}% Align table columns on decimal point
\usepackage{bm}% bold math
\usepackage{amsfonts}
\usepackage{mathrsfs}
\usepackage{color}
\usepackage{comment}

\def\bequ{\begin{equation}}
\def\eequ{\end{equation}}
\def\be{\begin{equation}}
\def\ee{\end{equation}}

\begin{document}

%\preprint{APS/123-QED}

\title{Maxwell quasinormal modes on a global monopole Schwarzschild-anti-de Sitter black hole with Robin boundary conditions}

\author{Mengjie Wang}
\email{mjwang@hunnu.edu.cn.}
%\email{Corresponding author: mjwang@hunnu.edu.cn.}
%
\author{Zhou Chen}
\author{Qiyuan Pan}
\email{panqiyuan@hunnu.edu.cn}
\author{Jiliang Jing}
\email{jljing@hunnu.edu.cn}
\affiliation{\vspace{2mm}
Department of Physics and Synergetic Innovation Center for Quantum Effects and Applications, Hunan Normal University,
Changsha, Hunan 410081, P.R. China \vspace{1mm}}
\date{\today}

\begin{abstract}
We generalize our previous studies on the Maxwell quasinormal modes around Schwarzschild-anti-de-Sitter black holes with Robin type vanishing energy flux boundary conditions, by adding a \textit{global monopole} on the background. We first formulate the Maxwell equations both in the Regge-Wheeler-Zerilli and in the Teukolsky formalisms and derive, based on the vanishing energy flux principle, \textit{two} boundary conditions in each formalism. The Maxwell equations are then solved analytically in pure anti-de Sitter spacetimes with a global monopole, and two \textit{different} normal modes are obtained due to the existence of the monopole parameter. In the small black hole and low frequency approximations, the Maxwell quasinormal modes are solved perturbatively on top of normal modes by using an asymptotic matching method, while beyond the aforementioned approximation, the Maxwell quasinormal modes are obtained numerically. We analyze the Maxwell quasinormal spectrum by varying the angular momentum quantum number $\ell$, the overtone number $N$, and in particular, the monopole parameter $8\pi\eta^2$. We show explicitly, through calculating quasinormal frequencies with both boundary conditions, that the global monopole produces the repulsive force.
\end{abstract}

\maketitle

%%%%%%%%%%%%%%%%%
\section{Introduction}
%%%%%%%%%%%%%%%%%%%%%%%%%%%%%%%%%%%%%%%%%%%%%%%%%%%%%%%%%%%%%%%%%%%%%%%%%%%%%%
Black hole quasinormal modes (QNMs), describing the characteristic oscillations of black holes, have attracted a lot of attention recently, see for example reviews~\cite{Chandrasekhar:1985kt,Kokkotas:1999bd,Berti:2009kk,Konoplya:2011qq} and references therein. Due to the existence of an event horizon, the black hole spacetimes are intrinsically dissipative so that quasinormal frequencies are complex in general and the imaginary part is associated with the timescale of the perturbation. QNMs play vital roles on various aspects, ranging from gravitational wave astronomy~\cite{Berti:2015itd,Barack:2018yly} to the application in the context of the anti--de Sitter/conformal field theory (AdS/CFT) correspondence~\cite{Maldacena:1997re,Gubser:1998bc,Witten:1998qj}.   

The AdS/CFT correspondence states that QNMs of a ($D+1$)-dimensional asymptotically AdS black hole or brane are poles of the retarded Green's function in the dual conformal field theory in $D$ dimensions at strong coupling. Horowitz and Hubeny first studied scalar QNMs on Schwarzschild-AdS black holes~\cite{Horowitz:1999jd} (see also~\cite{Chan:1996yk,Chan:1999sc}), and numerous works were then followed to explore QNMs of various spin fields on asymptotically AdS black holes, see for example~\cite{Wang:2000gsa,Wang:2000dt,Govindarajan:2000vq,Zhu:2001vi,Birmingham:2001hc,Cardoso:2001bb,Cardoso:2001vs,Moss:2001ga,Birmingham:2001pj,Konoplya:2002ky,Musiri:2003rv,Berti:2003ud,Jing:2005uy,Hertog:2004bb,Giammatteo:2004wp,Siopsis:2004as,Gutsche:2019blp,Abdalla:2019irr,Che:2019jvy,Lin:2019fte,Aragon:2020tvq,Chernicoff:2020kmf,Konoplya:2017zwo,Hendi:2018hdo,Gonzalez:2018xrq}.  

Mathematically QNMs are defined as eigenvalues of perturbation equations with physically relevant boundary conditions. Considering a lot of studies already performed in literatures, however, a generic boundary condition is still lacking. Recently, we have proposed the vanishing energy flux principle~\cite{Wang:2015goa,Wang:2016dek}, which may be applied both to the Regge-Wheeler-Zerilli and to the Teukolsky formalisms, and leads to two sets of Robin type boundary conditions and has been successfully employed to explore QNMs of the Maxwell~\cite{Wang:2015goa,Wang:2015fgp} and Dirac fields~\cite{Wang:2017fie,Wang:2019qja}. In this paper, we follow the same rationale and generalize our previous studies of the Maxwell QNMs on Schwarzschild-AdS black holes, by adding a \textit{global monopole} on the backgrounds.

The global monopoles, as a special class of topological defects, may be formed in the early universe through the spontaneous symmetry breaking of the global O(3) symmetry to U(1)~\cite{Kibble:1976sj,Vilenkin:1984ib}, according to the Grand Unified Theories. The gravitational properties of monopoles have been extensively studied, and an unusual property induced by global monopoles is that it possesses a solid deficit angle. This property makes black holes with a global monopole and without a global monopole topologically different, and thus leads to interesting physical consequences~\cite{Pan:2008xz,Chen:2005vq,Yu:2002st,Chen:2009vz,Piedra:2019ytw,Secuk:2019njc,Soroushfar:2020wch,Zhang:2014xha}.  

The purpose of this study is twofold. On one hand, we explore the impact of the global monopole on the Maxwell quasinormal spectrum on Schwarzschild-AdS black holes, by imposing vanishing energy flux boundary conditions. On the other hand, it is well known that, on spherically symmetric backgrounds, the Maxwell equations may be written either in the Regge-Wheeler-Zerilli or in the Teukolsky formalisms. As we argued before~\cite{Wang:2015goa}, by imposing vanishing energy flux boundary conditions, the Maxwell equations in both formalisms lead to the same quasinormal spectrum. Here we show explicitly, through calculating normal modes in both formalisms with vanishing energy flux boundary conditions, that it is \textit{indeed} the case, even if a global monopole is included.

The structure of this paper is organized as follows. In Section~\ref{seceq} we introduce the Schwarzschild-AdS black holes with a global monopole, and show the Maxwell equations both in the Regge-Wheeler-Zerilli and in the Teukolsky formalisms. In Section~\ref{secbc} we present the explicit boundary conditions, based on the vanishing energy flux principle, for both the Regge-Wheeler-Zerilli variable and the Teukolsky variable of the Maxwell field. We then perform an analytic matching calculation for small AdS black holes in Section~\ref{secana}, and a numeric calculation in Section~\ref{secnum}. Final remarks and conclusions are presented in the last section.

%%%%%%%%%%%%%%%%%%%%%%%%%%%%%%%%%%%%%
\section{background geometry and the field equations}
\label{seceq}
%%%%%%%%%%%%%%%%%%%%%%%%%%%%%%%%%%%%%
In this section, we first briefly review the background geometry we shall study, i.e. Schwarzschild-AdS black holes with a global monopole, and then present equations of motion for the Maxwell fields on the aforementioned backgrounds both in the Regge-Wheeler-Zerilli and in the Teukolsky formalisms.
  
\subsection{The line element}
We start by considering the following line element of a Schwarzschild-AdS black hole with a global monopole
\begin{equation}
ds^2=\dfrac{\Delta_r}{r^2}dt^2-\dfrac{r^2}{\Delta_r}dr^2-r^2\left(d\theta^2+\sin^2\theta d\varphi^2\right) \;,\label{metric}
\end{equation}
with the metric function
\begin{equation}
\Delta_r\equiv r^2\left(\tilde{\eta}^2+\frac{r^2}{L^2}\right)-2Mr\;,\label{metricfunc}
\end{equation}
where $L$ is the AdS radius, $M$ is the mass parameter. Here the dimensionless parameter $\tilde{\eta}^2$ is defined by
\begin{equation}
\tilde{\eta}^2\equiv 1-8\pi\eta^2\;,\label{monopole}
\end{equation}
where $\eta$ is the global monopole parameter, and the Schwarzschild-AdS spacetimes may be recovered when $8\pi\eta^2=0$. The Hawking temperature may be calculated, and one obtains
\begin{equation}
T_H=\dfrac{\kappa}{2\pi}=\dfrac{3r_+^2+\tilde{\eta}^2L^2}{4\pi r_+L^2}\;,\nonumber
\end{equation}
where $r_+$ is the event horizon determined by the non-zero real root of $\Delta_r(r_+)=0$, and where the mass parameter has been expressed in terms of $r_+$ as
\begin{equation}
M=\dfrac{r_+(\tilde{\eta}^2L^2+r_+^2)}{2L^2}\;.\nonumber
\end{equation}

By introducing the following coordinates transformation
\begin{equation}
\tilde{t}=\tilde{\eta}t\;,\;\;\;\;\;\;\tilde{r}=\dfrac{r}{\tilde{\eta}}\;,\label{ttrt}
\end{equation}
and a new mass paramter
\begin{equation}
\tilde{M}=\dfrac{M}{\tilde{\eta}^3}\;,\label{masst}
\end{equation}
Eq.~\eqref{metric} becomes
\begin{align}
ds^2=&\left(1-\frac{2\tilde{M}}{\tilde{r}}+\frac{\tilde{r}^2}{L^2}\right)d\tilde{t}^{\;2}-\left(1-\frac{2\tilde{M}}{\tilde{r}}+\frac{\tilde{r}^2}{L^2}\right)^{-1}d\tilde{r}^2\nonumber\\&-\tilde{\eta}^2\tilde{r}^2\left(d\theta^2+\sin^2\theta d\varphi^2\right) \;.\label{metrict}
\end{align}
Now it becomes clear that the global monopole introduces a solid deficit angle, so that the solid angle of the above spacetime is $4\pi\tilde{\eta}^2$. 

%%%%%%
\subsection{Equations of motion in the Regge-Wheeler-Zerilli formalism}
%%%%%%
In a spherically symmetric background, one may obtain variable separated and decoupled Maxwell equations by using the Regge-Wheeler-Zerilli method~\cite{Regge:1957td, Zerilli:1970se}. For that purpose, we start from the Maxwell equations
\begin{equation}
\nabla_{\nu}F^{\mu\nu}=0\;,\label{Maxwelleq}
\end{equation}
where the field strength tensor is defined as $F_{\mu\nu}=\partial_{\mu}A_{\nu}-\partial_{\nu}A_{\mu}$. 
We then expand the vector potential $A_\mu$ in terms of the scalar and vector spherical harmonics~\cite{Ruffini:1973}
\begin{equation}
A_{\mu}=\sum_{\ell, m}\left(\left[\begin{array}{c} 0 \\
0 \\
a^{\ell m}(t,r) \boldsymbol {S}_{\ell m}\end{array}\right]+\left[
\begin{array}{c}j^{\ell m}(t,r)Y_{\ell m}  \\
h^{\ell m}(t,r)Y_{\ell m} \\
k^{\ell m}(t,r)\boldsymbol {Y}_{\ell m}  
\end{array}\right]\right)\;,\label{Vpotential}
\end{equation}
with the definition of the vector spherical harmonics
\begin{equation}
\boldsymbol {S}_{\ell m}=
\left(\begin{array}{c} \frac{1}{\sin \theta} \partial_{\varphi}Y_{\ell m}  \\
-\sin \theta \partial_{\theta}Y_{\ell m}\end{array}\right)\;,\;\;\;
\boldsymbol {Y}_{\ell m}=
\left(\begin{array}{c} \partial_{\theta}Y_{\ell m}  \\
\partial_{\varphi}Y_{\ell m}\end{array}\right)\;,\nonumber
\end{equation}
where $Y_{\ell m}$ are the scalar spherical harmonics, $m$ is the azimuthal number, and $\ell$ is the angular momentum quantum number. Note that the first term in the right hand side of Eq.~\eqref{Vpotential} has parity $(-1)^{\ell+1}$ while the second term has parity $(-1)^\ell$, and we shall call the former (latter) the axial (polar) modes. By substituting Eq.~\eqref{Vpotential} into Eq.~\eqref{Maxwelleq} with the assumption
\begin{align}
a^{\ell m}(t,r)=e^{-i\omega t}a^{\ell m}(r)\;,\;\;\;j^{\ell m}(t,r)=e^{-i\omega t}j^{\ell m}(r)\;,\nonumber\\
h^{\ell m}(t,r)=e^{-i\omega t}h^{\ell m}(r)\;,\;\;\;k^{\ell m}(t,r)=e^{-i\omega t}k^{\ell m}(r)\;,\nonumber
\end{align}
one obtains the Schrodinger-like radial wave equation 
\begin{equation}
\left(\frac{d^2}{dr_{*}^2}+\omega^2-\ell(\ell+1)\dfrac{\Delta_r}{r^4}\right)\Psi(r)=0\;,\label{RWZeq}
\end{equation}
where the tortoise coordinate is defined as
\begin{equation}
\dfrac{dr_*}{dr}=\dfrac{r^2}{\Delta_r}\;,\label{tortoisecoor}
\end{equation}
with $\Psi(r)=a^{\ell m}(r)$ for axial modes, and 
\begin{equation}
\Psi(r)=\dfrac{r^2}{\ell(\ell+1)}\left(-i\omega h^{\ell m}(r)-\dfrac{dj^{\ell m}(r)}{dr}\right)\;,\nonumber
\end{equation}
for polar modes.

%%%%%%
\subsection{Equations of motion in the Teukolsky formalism}
%%%%%%
Equations of motion of the Maxwell fields may be also derived within the Teukolsky formalism~\cite{Teukolsky:1973ha}. This approach is based on the Newmann-Penrose algorithm~\cite{Newman:1961qr}, and is particularly relevant to study linear perturbations of the massless spin fields on rotating black hole backgrounds. In this subsection we outline the radial equations, which may be obtained following the procedures presented in~\cite{Khanal:1983vb}.

The radial equation is
\begin{equation}
\Delta_r^{-s}\dfrac{d}{dr}\left(\Delta_r^{s+1}\dfrac{d R_{s}(r)}{dr}\right)+H(r)R_{s}(r)=0\;,\label{Teukeq}
\end{equation}
with
\begin{eqnarray}
H(r)=\dfrac{K_r^2-i s K_r \Delta_r^\prime}{\Delta_r}+2is K_r^\prime +\dfrac{s+|s|}{2}\Delta_r^{\prime\prime}
-\lambda\;,\nonumber
\end{eqnarray}
where $K_r=\omega r^2$, $\lambda=\ell(\ell+1)$ and the spin parameter is $s=\pm1$.

%%%%%%%%%%%%%%%%%%%%%%%%%%%%%%%%%%%%%
\section{boundary conditions}
\label{secbc}
%%%%%%%%%%%%%%%%%%%%%%%%%%%%%%%%%%%%%
In order to solve the radial equations, given by Eqs.~\eqref{RWZeq} and~\eqref{Teukeq}, one has to impose physically relevant boundary conditions, both at the horizon and at infinity. At the horizon, we impose the commonly used ingoing wave boundary conditions. At infinity, we impose \textit{the vanishing energy flux principle}, proposed in~\cite{Wang:2015goa} (see also~\cite{Wang:2016dek,Wang:2016zci}), which have already been employed to study the Maxwell~\cite{Wang:2015goa,Wang:2015fgp} and the Dirac~\cite{Wang:2017fie,Wang:2019qja} QNMs on asymptotically AdS spacetimes. Based on this principle, in the following we derive explicit boundary conditions for Eqs.~\eqref{RWZeq} and~\eqref{Teukeq}, which are obtained in the Regge-Wheeler-Zerilli and in the Teukolsky formalisms respectively, and we will show both equations with the corresponding boundary conditions lead to the same spectrum in the next section.

%%%%%%
\subsection{Boundary conditions in the Regge-Wheeler-Zerilli formalism}
%%%%%%
We start from the energy-momentum tensor of the Maxwell field, which is given by
\begin{equation}
T_{\mu \nu}=F_{\mu\sigma}F^\sigma_{\;\;\;\nu}+\dfrac{1}{4}g_{\mu\nu}F^2\;.\label{EMTensor}
\end{equation}
Then the spatial part of the radial energy flux may be calculated as 
\begin{equation}
\mathcal{F}|_r\propto\dfrac{\Delta_r}{r^2}\Psi(r)\Psi^\prime(r)\;,\label{RWZbc1}
\end{equation}
where $\prime$ denotes the derivative with respect to $r$.
By expanding Eq.~\eqref{RWZeq} asymptotically as
\begin{equation}
\Psi\sim a_{0}+\frac{a_{1}}{r}+\mathcal{O}\left(\frac{1}{r^2}\right)\;,\label{RWZasysol}
\end{equation}
Eq.~\eqref{RWZbc1} becomes
\begin{equation}
\mathcal{F}|_{r,\infty}\propto a_0a_1\;.\nonumber
\end{equation}
Then the vanishing energy flux principle, i.e. $\mathcal{F}|_{r,\infty}=0$, leads to 
\begin{align}
a_0&=0\;,\label{RWZbc2-1}\\
a_1&=0\;.\label{RWZbc2-2}
\end{align}

\subsection{Boundary conditions in the Teukolsky formalism}
The explicit boundary conditions for the Teukolsky variables of the Maxwell fields on a global monopole Schwarzschild-AdS black hole can be derived directly, following the similar prescriptions described in~\cite{Wang:2015goa,Wang:2016zci}. Since the monopole parameter does not alter the asymptotic structure of AdS spacetimes, one may get exactly the same boundary conditions as to the Schwarzschild-AdS case, and the results are listed in the following.

To be specific, we focus on the boundary conditions for $R_{-1}$. From Eq.~\eqref{Teukeq} one obtains the asymptotic behavior of $R_{-1}$ as
\begin{equation}
R_{-1} \sim \;\alpha^{-} r+\beta^{-}+\mathcal{O}(r^{-1})\;,\label{asysol}
\end{equation}
and the vanishing energy flux principle leads to~\cite{Wang:2015goa,Wang:2016zci}
\begin{align}
&\dfrac{\alpha^{-}}{\beta^{-}}=\dfrac{i}{\omega L^2}\;,\label{Teubc1}
\\
&\dfrac{\alpha^{-}}{\beta^{-}}=\dfrac{i\omega}{-\ell(\ell+1)+\omega^2L^2}\; .\label{Teubc2}
\end{align}

%%%%%%%%%%%%%%%%%%%%%%%%%%%%%%%%%%%%%
\section{Analytics}
\label{secana}
%%%%%%%%%%%%%%%%%%%%%%%%%%%%%%%%%%%%%
%%%%%%
\subsection{Normal modes}
%%%%%%
The normal modes of the Maxwell fields on an empty AdS background with a global monopole are calculated \textit{analytically} in this subsection, both in the Regge--Wheeler--Zerilli and in the Teukolsky formalisms, by solving Eq.~\eqref{RWZeq} with boundary conditions~\eqref{RWZbc2-1}  and~\eqref{RWZbc2-2}, and Eq.~\eqref{Teukeq} with boundary conditions~\eqref{Teubc1}  and~\eqref{Teubc2}. These calculations provide a concrete example to show \textit{explicitly} that, \textit{vanishing energy flux} is a generic principle, which can be applied to both formalisms and leads to the same spectrum. 

\subsubsection{Normal modes in the Regge--Wheeler--Zerilli formalism}
In a pure AdS spacetime with a global monopole ($M=0$), the metric function becomes
\begin{equation}
\Delta_r=r^2\left(\tilde{\eta}^2+\dfrac{r^2}{L^2}\right)\;,\nonumber
\end{equation}
then the radial equation~\eqref{RWZeq} can be solved, and one obtains
%%%%%%%%%%
\begin{align}
&\Psi(r)=r^{\tilde{\ell}+1}\left(r^2+\tilde{L}^2\right)^{-\frac{\tilde{\omega}\tilde{L}}{2}}\Big[c_1F\Big(\frac{1+\tilde{\ell}-\tilde{\omega}\tilde{L}}{2},\Big.\Big.\nonumber \\ & \Big.\Big.\frac{2+\tilde{\ell}-\tilde{\omega}\tilde{L}}{2},\frac{3}{2}+\tilde{\ell};-\frac{r^2}{\tilde{L}^2}\Big)-c_2e^{-2i\pi\tilde{\ell}}\Big(\frac{\tilde{L}}{r}\Big)^{2\tilde{\ell}+1}\Big.\nonumber \\ & \Big.F\Big(-\frac{\tilde{\ell}+\tilde{\omega}\tilde{L}}{2},\frac{1-\tilde{\ell}-\tilde{\omega}\tilde{L}}{2},\frac{1}{2}-\tilde{\ell};-\frac{r^2}{\tilde{L}^2}\Big)\Big]\;.\label{AdSRWZeq}
\end{align}
%%%%%%%%%%
Here $c_1$, $c_2$ are two integration constants with dimension of inverse length, $F(a,b,c,z)$ is the hypergeometric function, and
\begin{equation}
\tilde{\ell}=\dfrac{1}{2}\Big(-1+\dfrac{\sqrt{4\ell^2+4\ell+\tilde{\eta}^2}}{\tilde{\eta}}\Big)\;,\;
\tilde{L}=\tilde{\eta}L\;,\;\tilde{\omega}=\dfrac{\omega}{\tilde{\eta}^2}\;,\label{pararelation}
\end{equation}
where $\ell=1,2,3,\cdot\cdot\cdot$. By expanding Eq.~\eqref{AdSRWZeq} at large $r$, we get relations between $c_1$ and $c_2$, i.e.
\begin{equation}
\dfrac{c_2}{c_1}=e^{2i\pi\tilde{\ell}}\dfrac{\Gamma\left(\frac{3}{2}+\tilde{\ell}\right)\Gamma\left(\frac{1-\tilde{\ell}-\tilde{\omega}\tilde{L}}{2}\right)\Gamma\left(\frac{1-\tilde{\ell}+\tilde{\omega}\tilde{L}}{2}\right)}{\Gamma\left(\frac{1}{2}-\tilde{\ell}\right)\Gamma\left(\frac{2+\tilde{\ell}-\tilde{\omega}\tilde{L}}{2}\right)\Gamma\left(\frac{2+\tilde{\ell}+\tilde{\omega}\tilde{L}}{2}\right)}\;,\label{RWZrela1}
\end{equation}
which corresponds to the first boundary condition given by Eq.~\eqref{RWZbc2-1}, and
\begin{equation}
\dfrac{c_2}{c_1}=e^{2i\pi\tilde{\ell}}\frac{\Gamma\left(\frac{3}{2}+\tilde{\ell}\right)\Gamma\left(\frac{-\tilde{\ell}-\tilde{\omega}\tilde{L}}{2}\right)\Gamma\left(\frac{-\tilde{\ell}+\tilde{\omega}\tilde{L}}{2}\right)}{\Gamma\left(\frac{1}{2}-\tilde{\ell}\right)\Gamma\left(\frac{1+\tilde{\ell}-\tilde{\omega}\tilde{L}}{2}\right)\Gamma\left(\frac{1+\tilde{\ell}+\tilde{\omega}\tilde{L}}{2}\right)}\;,\label{RWZrela2}
\end{equation}
which corresponds to the second boundary condition given by Eq.~\eqref{RWZbc2-2}.
Then by expanding Eq.~\eqref{AdSRWZeq} at small $r$
\begin{equation}
\Psi(r)\sim\;c_1r^{1+\tilde{\ell}}\tilde{L}^{-\tilde{\omega}\tilde{L}}-c_2e^{-2i\pi\tilde{\ell}}r^{-\tilde{\ell}}\tilde{L}^{1+2\tilde{\ell}-\tilde{\omega}\tilde{L}}\;,
\end{equation}
we shall set $c_2=0$ to get a regular solution at the origin. This condition leads to two sets of normal modes
\begin{equation}
\Gamma\Big(\frac{2+\tilde{\ell}-\tilde{\omega}\tilde{L}}{2}\Big)=-N\;\Rightarrow\;\tilde{\omega}_{1, N} \tilde{L}=2N+\tilde{\ell}+2\;,\label{normal1}
\end{equation}
from Eq.~\eqref{RWZrela1}, and
\begin{equation}
\Gamma\Big(\frac{1+\tilde{\ell}-\tilde{\omega}\tilde{L}}{2}\Big)=-N\;\Rightarrow\;\tilde{\omega}_{2, N} \tilde{L}=2N+\tilde{\ell}+1\;,\label{normal2}
\end{equation}
from Eq.~\eqref{RWZrela2}, where $N=0,1,2,\cdot\cdot\cdot$. The above two normal modes, by noticing that $\tilde{\ell}$ is not an integer anymore, are \textit{different}. This is an interesting observation, since for the case without a global monopole, the two sets of the Maxwell normal modes are isospectral up to one mode~\cite{Wang:2015goa}. 

%%%%%%%%%%
\subsubsection{Normal modes in the Teukolsky formalism}
%%%%%%%%%%
In this case the radial Teukolsky equation~\eqref{Teukeq} becomes 
\begin{align}
&\Delta_rR_{-1}''(r)+\left(\dfrac{K_r^2+i K_r \Delta_r^\prime}{\Delta_r}-2iK_r^\prime
-\ell(\ell+1)\right)R_{-1}(r)\nonumber\\
&=0\;,\label{Teufareq1}
\end{align}
with
\begin{equation}
\Delta_r= r^2 \Big(\tilde{\eta}^2+\dfrac{r^2}{L^2}\Big)\;,\;\;\;\;\;\;K_r=\omega r^2 .\nonumber
\end{equation}
The general solution for Eq.~\eqref{Teufareq1} is 
\begin{align}
&R_{-1}=r^{\tilde{\ell}+1}(r-i\tilde{L})^{\frac{\tilde{\omega} \tilde{L}}{2}}(r+i\tilde{L})^{-\tilde{\ell}-\frac{\tilde{\omega}\tilde{L}}{2}}\Big[c_3F\Big(\tilde{\ell},\tilde{\ell}+1\Big.\Big.\nonumber \\ & \Big.\Big.+\tilde{\omega}\tilde{L},2\tilde{\ell}+2;\dfrac{2r}{r+i\tilde{L}}\Big)+c_4(-2)^{-2\tilde{\ell}-1}\Big(1+\dfrac{i\tilde{L}}{r}\Big)^{2\tilde{\ell}+1}\Big.\nonumber \\ & \Big.F\Big(-\tilde{\ell}-1,-\tilde{\ell}+\tilde{\omega}\tilde{L},-2\tilde{\ell};\dfrac{2r}{r+i\tilde{L}}\Big)\Big]\;,\label{Teufarsol}
\end{align}
where $F(a,b,c;z)$ is again the hypergeometric function, $c_3$ and $c_4$ are two integration constants with dimension of inverse length. These two constants are related to each other by the boundary conditions through expanding Eq.~\eqref{Teufarsol} at large $r$:
\begin{itemize}
\item[$\bullet$] By imposing the first boundary condition given in Eq.~\eqref{Teubc1}, one gets a first relation between $c_3$ and $c_4$
\begin{equation}
\dfrac{c_4}{c_3}=(-2)^{1+2\tilde{\ell}}\dfrac{\tilde{\ell}}{1+\tilde{\ell}}\dfrac{F(1+\tilde{\ell},1+\tilde{\ell}+\tilde{\omega}\tilde{L},2+2\tilde{\ell};2)}{F(-\tilde{\ell},-\tilde{\ell}+\tilde{\omega} \tilde{L},-2\tilde{\ell};2)}\;.\label{Teuc1c2bc1}
\end{equation}
\item[$\bullet$] By imposing the second boundary condition given in Eq.~\eqref{Teubc2}, on the other hand, one gets a second relation between $c_3$ and $c_4$
\begin{equation}
\dfrac{c_4}{c_3}=(-2)^{1+2\tilde{\ell}}\dfrac{\tilde{\ell}}{\tilde{\ell}+1}\dfrac{\mathcal{A}_1}{\mathcal{A}_2}\;,\label{Teuc1c2bc2}
\end{equation}
where
\begin{align}
\mathcal{A}_1=&(1+\tilde{\ell}) F(\tilde{\ell},1+\tilde{\ell}+\tilde{\omega}\tilde{L},2+2\tilde{\ell};2)\nonumber\\&+\tilde{\omega}\tilde{L} F(1+\tilde{\ell},1+\tilde{\ell}+\tilde{\omega}\tilde{L},2+2\tilde{\ell};2)\;,\nonumber\\
\mathcal{A}_2=&\tilde{\ell} F(1-\tilde{\ell},-\tilde{\ell}+\tilde{\omega}\tilde{L},-2\tilde{\ell};2)\nonumber\\&-\tilde{\omega}\tilde{L} F(-\tilde{\ell},-\tilde{\ell}+\tilde{\omega}\tilde{L},-2\tilde{\ell};2)\;.\label{expA}
\end{align}
\end{itemize}
Then from the small $r$ behavior of Eq.~\eqref{Teufarsol}
\begin{equation}
R_{-1}\sim c_3e^{i\pi\tilde{\ell}}2^{1+2\tilde{\ell}}\tilde{L}^{-2\tilde{\ell}}r^{\tilde{\ell}+1}+c_4\dfrac{-i \tilde{L}}{r^{\tilde{\ell}}}\;,\label{farsolnear}
\end{equation}
we have to set $c_4=0$ in order to get a regular solution of $R_{-1}$ at the origin. This regularity condition picks the normal modes, from Eqs.~\eqref{Teuc1c2bc1} and~\eqref{Teuc1c2bc2}:
\begin{eqnarray}
&&F(1+\tilde{\ell},1+\tilde{\ell}+\tilde{\omega}\tilde{L},2+2\tilde{\ell};2)=0\;,\nonumber\\
&&\Rightarrow\;\;\tilde{\omega}_{1,N}\tilde{L}=2N+\tilde{\ell}+2\;,\label{Teuknormalmode1}\\
&&\mathcal{A}_1=0\;,\nonumber\\&&\Rightarrow\;\;\tilde{\omega}_{2,N}\tilde{L}=2N+\tilde{\ell}+1\;,\label{Teuknormalmode2}
\end{eqnarray}
where again $N=0,1,2,\cdot\cdot\cdot$, and two sets of normal modes are \textit{different}. One may observe that normal modes obtained in the Teukolsky formalism, given in Eqs.~\eqref{Teuknormalmode1} and~\eqref{Teuknormalmode2}, are exactly the same with the counterpart obtained in the Regge-Wheeler-Zerilli formalism, given in Eqs.~\eqref{normal1} and~\eqref{normal2}, which indicates the equivalence of the two formalisms and the universality of the vanishing energy flux boundary conditions.

%%%%%%
\subsection{Quasinormal modes for small black holes}
%%%%%%
In this subsection, we perform an analytic calculation of quasinormal frequencies for the Maxwell fields on a small Schwarzschild-AdS black hole with a global monopole, by using an asymptotic matching method. Note that for this case the analytic calculation is only applicable to the Teukolsky formalism.

%%%%%%
\subsubsection{Near region}
%%%%%%
In the near region, and with small black hole approximation $r_+\ll \tilde{L}$, Eq.~\eqref{Teukeq} becomes
\begin{equation}
\left(\Delta_r\dfrac{d^2}{dr^2}+\dfrac{\tilde{\eta}^4r_{+}^2\bar{\omega}}{\Delta_{r}}-\ell(\ell+1)\right)R_{-1}(r)=0\;,\label{Teumatch1}
\end{equation}
with
\begin{equation}
\bar{\omega}=\Big(\tilde{\omega}r_++\dfrac{i}{2}\Big)^2+\frac{1}{4}\;,\;\;\;\Delta_r=\tilde{\eta}^2r(r-r_+)\;,\label{omegabar}
\end{equation}
where $\tilde{\omega}$ is defined in Eq.~\eqref{pararelation}. By defining a new dimensionless variable
\begin{equation}
z\equiv1-\dfrac{r_+}{r}\;,\nonumber
\end{equation}
it is convenient to transform Eq.~\eqref{Teumatch1} into
\begin{align}
&z(1-z)\dfrac{d^2R_{-1}}{dz^2}-2z\dfrac{dR_{-1}}{dz}+\Big(\dfrac{\bar{\omega}(1-z)}{z}-\dfrac{\tilde{\ell}(\tilde{\ell}+1)}{(1-z)}\Big)R_{-1}\nonumber\\
&=0\;,\label{Teumatch2}
\end{align}
where $R_{-1}\equiv R_{-1}(z)$, and $\tilde{\ell}$ is defined in Eq.~\eqref{pararelation}. The above equation can be solved in terms of the hypergeometric function
\begin{equation}
R_{-1}\sim z^{1-i\tilde{\omega}r_+} (1-z)^{\tilde{\ell}}F(\tilde{\ell}+1, \tilde{\ell}+2-2i\tilde{\omega}r_+, 2-2i\tilde{\omega}r_+; z)\;,\label{matchingsolnear}
\end{equation}
where an ingoing boundary condition has been imposed. In order to match the far region solution, here we shall further expand the near region solution, given in Eq.~\eqref{matchingsolnear}, at large $r$. To do so, we take the $z\rightarrow1$ limit and use the property of the hypergeometric function~\cite{abramowitz+stegun}, then obtain
\begin{equation}
R_{-1}\sim \Gamma(2-2i\tilde{\omega}r_+) \left[\dfrac{R^{\rm near}_{-1,1/r}}{r^{\tilde{\ell}}}+R^{\rm near}_{-1,r}r^{\tilde{\ell}+1}\right]\;,\label{Teuknearfar}
\end{equation}
where
\begin{align}
&R^{\rm near}_{-1,1/r}=\dfrac{\Gamma(-2\tilde{\ell}-1)r_+^{\tilde{\ell}}}{\Gamma(1-\tilde{\ell}-2i\tilde{\omega}r_+)\Gamma(-\tilde{\ell})}\;,\nonumber\\
&R^{\rm near}_{-1,r}=\dfrac{\Gamma(2\tilde{\ell}+1)r_+^{-\tilde{\ell}-1}}{\Gamma(\tilde{\ell}+1)\Gamma(\tilde{\ell}+2-2i\tilde{\omega}r_+)}\;.\label{TeuknearfarCoeff}
\end{align}

%%%%%%
\subsubsection{Far region}
%%%%%%
In the far region, the black hole effects may be neglected, and the solution is given by Eq.~\eqref{Teufarsol}. In order to match this solution with the near region solution, we shall expand Eq.~\eqref{Teufarsol} at small $r$, then obtain
\begin{equation}
R_{-1}\sim\dfrac{R^{\rm far}_{-1,1/r}}{r^{\tilde{\ell}}}+R^{\rm far}_{-1,r}r^{\tilde{\ell}+1}\;,\label{Teukfarnear}
\end{equation}
with
\begin{equation}
R^{\rm far}_{-1,1/r}\equiv -i\tilde{L}c_4\;,\;\;\;
R^{\rm far}_{-1,r}\equiv2^{1+2\tilde{\ell}}e^{i\pi\tilde{\ell}}\tilde{L}^{-2\tilde{\ell}}c_3\;,\label{TeukfarnearCoeff}
\end{equation}
where the constants $c_3$ and $c_4$ are related with each other by Eqs.~\eqref{Teuc1c2bc1} and~\eqref{Teuc1c2bc2}, corresponding to the first and second boundary conditions.

%%%%%%
\subsubsection{Overlap region}
%%%%%%
In the overlap region the solutions, obtained in the near region given by Eq.~\eqref{Teuknearfar} and in the far region given by Eq.~\eqref{Teukfarnear}, are the same up to a constant. Then one may impose the matching condition, $R^{\rm near}_{-1, r}R^{\rm far}_{-1, 1/r}=R^{\rm far}_{-1, r}R^{\rm near}_{-1, 1/r}$, which gives
\begin{align}
&\dfrac{\Gamma(-2\tilde{\ell}-1)}{\Gamma(-\tilde{\ell})}\dfrac{\Gamma(\tilde{\ell}+1)}{\Gamma(2\tilde{\ell}+1)}\dfrac{\Gamma(\tilde{\ell}+2-2i\tilde{\omega}r_+)}{\Gamma(-\tilde{\ell}+1-2i\tilde{\omega}r_+)}\left(\dfrac{r_+}{\tilde{L}}\right)^{2\tilde{\ell}+1}\nonumber\\
&=\left(\dfrac{-i}{2}\right)^{1+2\tilde{\ell}}\dfrac{c_4}{c_3}\;.\label{Teukrel}
\end{align}
By imposing the \textit{first} boundary condition and using the corresponding relation between $c_3$ and $c_4$ given by Eq.~\eqref{Teuc1c2bc1}, one obtains
\begin{align}
&\dfrac{\Gamma(-2\tilde{\ell}-1)}{\Gamma(-\tilde{\ell})}\dfrac{\Gamma(\tilde{\ell}+1)}{\Gamma(2\tilde{\ell}+1)}\dfrac{\Gamma(\tilde{\ell}+2-2i\tilde{\omega}r_+)}{\Gamma(-\tilde{\ell}+1-2i\tilde{\omega}r_+)}\left(\dfrac{r_+}{\tilde{L}}\right)^{2\tilde{\ell}+1}\nonumber\\
&=i^{1+2\tilde{\ell}}\dfrac{\tilde{\ell}}{1+\tilde{\ell}}\dfrac{F(1+\tilde{\ell},1+\tilde{\ell}+\tilde{\omega}\tilde{L},2+2\tilde{\ell};2)}{F(-\tilde{\ell},-\tilde{\ell}+\tilde{\omega} \tilde{L},-2\tilde{\ell};2)}\;,\label{Teukqnm1}
\end{align}
while by imposing the \textit{second} boundary condition and using the corresponding relation between $c_3$ and $c_4$ given by Eq.~\eqref{Teuc1c2bc2}, one obtains
\begin{align}
&\dfrac{\Gamma(-2\tilde{\ell}-1)}{\Gamma(-\tilde{\ell})}\dfrac{\Gamma(\tilde{\ell}+1)}{\Gamma(2\tilde{\ell}+1)}\dfrac{\Gamma(\tilde{\ell}+2-2i\tilde{\omega}r_+)}{\Gamma(-\tilde{\ell}+1-2i\tilde{\omega}r_+)}\left(\dfrac{r_+}{\tilde{L}}\right)^{2\tilde{\ell}+1}\nonumber\\
&=i^{1+2\tilde{\ell}}\dfrac{\tilde{\ell}}{1+\tilde{\ell}}\dfrac{\mathcal{A}_1}{\mathcal{A}_2}\;,\label{Teukqnm2}
\end{align}
where $\mathcal{A}_1$ and $\mathcal{A}_2$ are given by Eq.~\eqref{expA}. 

For a small black hole ($r_+\ll \tilde{L}$), at the leading order of $r_+/\tilde{L}$, the left terms in Eqs.~\eqref{Teukqnm1} and~\eqref{Teukqnm2} vanish, and then we shall require the right terms in both equations to vanish as well. These conditions lead to two sets of normal modes, given by Eqs.~\eqref{Teuknormalmode1} and~\eqref{Teuknormalmode2}. Then QNMs of small black holes may be obtained perturbatively by solving Eqs.~\eqref{Teukqnm1} and~\eqref{Teukqnm2}, on top of normal modes. To achieve this goal, we expand the frequency
\begin{equation}
\tilde{\omega}_j\tilde{L}=\tilde{\omega}_{j, N}\tilde{L}+i\delta_j\;,\label{Teukexpan}
\end{equation}
where $j=1, 2$, and $\tilde{\omega}_{j, N}$ refer to normal modes. Here $\delta_j$ is complex in general, and its real part, i.e. $\Re(\delta_j)$, reflects damping rate of a black hole. The general expression of $\delta_j$, which is usually messy and lengthy but can be derived straightforwardly by substituting Eq.~\eqref{Teukexpan} into Eqs.~\eqref{Teukqnm1} and~\eqref{Teukqnm2}.

%%%%%%%%%%%%%%%%%%%%%%%%%%%%%%%%%%%%%
\section{Numeric reuslts}
\label{secnum}
%%%%%%%%%%%%%%%%%%%%%%%%%%%%%%%%%%%%%
Beyond the regime where the asymptotic matching method is valid, one has to look for black hole quasinormal spectrum by resorting to numerics. In this part, we utilize a numeric pseudospectral method, adapted from our previous works~\cite{Wang:2019qja,Wang:2021upj}, to solve the Maxwell equations given in the Regge-Wheeler-Zerilli formalism~\eqref{RWZeq} with the corresponding boundary conditions given by Eqs.~\eqref{RWZbc2-1} and~\eqref{RWZbc2-2}.~\footnote{Note that, as we have checked, the same spectrum may be also obtained by solving the Teukolsky equation~\eqref{Teukeq} with the corresponding boundary conditions given by Eqs.~\eqref{Teubc1} and~\eqref{Teubc2}.} 

Before we introduce the pseudospectral method, here goes a few comments on the dimensionless form of Eq.~\eqref{RWZeq} which is essential for numeric calculations. For the case we considered in this paper, one may either take the unit of $L$ or take the unit of $\tilde{L}$ (with the definition given in Eq.~\eqref{pararelation}). For the former choice, Eq.~\eqref{RWZeq} may be written as
\begin{equation}
\left[\frac{\Delta_r}{r^2}\frac{d}{dr}\left(\frac{\Delta_r}{r^2}\frac{d}{dr}\right)+\omega^2L^2-\ell(\ell+1)\dfrac{\Delta_r}{r^4}\right]\Psi(r)=0\;,\label{RWZeq2}
\end{equation}
where $r$ is an abbreviation of $\tfrac{r}{L}$ so that it is dimensionless, and
\begin{equation}
\frac{\Delta_r}{r^2}=\frac{r-r_+}{r}\left(\tilde{\eta}^2+r^2+r_+r+r_+^2\right)\;,
\end{equation}
where $r_+$ is a dimensionless event horizon.

By taking the unit of $\tilde{L}$, Eq.~\eqref{RWZeq} becomes
\begin{equation}
\left[g(r)\frac{d}{dr}\left(g(r)\frac{d}{dr}\right)+\tilde{\omega}^2\tilde{L}^2-\tilde{\ell}(\tilde{\ell}+1)\dfrac{g(r)}{r^2}\right]\Psi(r)=0\;,\label{RWZeq3}
\end{equation}
with
\begin{equation}
g(r)=\frac{r-r_+}{r}\left(1+r^2+r_+r+r_+^2\right)\;,\label{eqg}
\end{equation}
where $r$ is an abbreviation of dimensionless radial coordinate $\tfrac{r}{\tilde{L}}$, $\tilde{\omega}$, $\tilde{L}$ and $\tilde{\ell}$ are given in Eq.~\eqref{pararelation}. 

By noticing that $g(r)$ has the same form with the metric of Schwarzschild-AdS, so Eq.~\eqref{RWZeq3} is exactly the same with the Maxwell equation on Schwarzschild-AdS, by replacing $\ell$ with $\tilde{\ell}$. This is also the Maxwell equation one may obtain by starting from the metric given by Eq.~\eqref{metrict}. Therefore, in our numeric calculations, we take the unit of $\tilde{L}$ and set $\tilde{L}=1$, and calculate the frequencies $\tilde{\omega}$. As we have checked, in the unit of $\tilde{L}$, the quasinormal frequencies have the uniform behaviors for various values of $r_+$, $\ell$ and $N$, and which is consistent with the physical picture that the global monopole produces the repulsive force. 

In order to employ a pseudospectral method conveniently, we first transform Eq.~\eqref{RWZeq}, which is a quadratic eigenvalue problem, into a linear eigenvalue problem, by
\begin{equation}
\Psi=e^{-i\omega r_\ast}\phi\;,\label{spectraltrans}
\end{equation}
where the tortoise coordinate $r_\ast$ is still defined in Eq.~\eqref{tortoisecoor}. Then changing the coordinate from $r$ to $z$ through
\begin{equation}
z=1-\dfrac{2r_+}{r}\;,\label{rtoz}
\end{equation}
which brings the integration domain from $r\in[r_+,\infty]$ to $z\in[-1,+1]$, and discretizing the $z$ coordinate according to the Chebyshev points
\begin{equation}
z_j=\cos\left(\dfrac{j\pi}{n}\right)\;,\;\;\;\;\;\;j=0,1,...,n\;,\label{spectralpoints}
\end{equation}
where $n$ denotes the number of grid points, Eq.~\eqref{RWZeq} turns into an algebraic equation
\begin{equation}
(M_0+\tilde{\omega} M_1)\phi(z)=0\;.\label{spectraleq2}
\end{equation}
Here $M_0$ and $M_1$ are matrices, which may be constructed straightforwardly by discretizing Eq.~\eqref{RWZeq} in terms of the Chebyshev points and Chebyshev differential matrices~\cite{trefethen2000spectral}.   

Boundary conditions associated to $\phi(z)$, may be derived from the transformation given by Eq.~\eqref{spectraltrans}. At the horizon, since an ingoing wave boundary condition is satisfied automatically, we simply impose a regular boundary condition for $\phi(z)$. At infinity, from Eqs.~\eqref{spectraltrans} and~\eqref{RWZasysol}, one obtains
\begin{equation}
\phi(z)=0\;,\label{spectralbc1}
\end{equation}
corresponding to the condition given in Eq.~\eqref{RWZbc2-1}, and
\begin{equation}
\dfrac{\phi^\prime(z)}{\phi(z)}=\dfrac{i\tilde{\omega}}{2r_+}\;,\label{spectralbc2}
\end{equation}
corresponding to the condition given in Eq.~\eqref{RWZbc2-2}.

\bigskip

One should note that we use $\tilde{\omega}_1$ ($\tilde{\omega}_2$) to represent the quasinormal frequency corresponding to the first (second) boundary conditions. A few selected data are presented below to demonstrate, in particular, the impact of global monopole on the spectrum. Also note that in our numeric calculations we focus on black holes with size $r_+\le1$ since in this regime the monopole effects are more relevant.~\footnote{Moreover, for large AdS black holes, the Maxwell spectrum may bifurcate, which has been explored in detail in our previous paper~\cite{Wang:2021upj}. }

%%%%%%%%%%%%%%%%%%%%%%%%%%%%%%%%%%%%%%%%%%%%%
\begin{table}
\caption{\label{EMmonopole} Quasinormal frequencies of the Maxwell fields on global monopole-Schwarzchild-AdS black holes with $8\pi\eta^2=0.1$, $N=0$, and for different black hole sizes $r_+$ with two different boundary conditions.}
\begin{ruledtabular}
\begin{tabular}{ l l l }
$r_+$ & $\tilde{\omega}_1 (\ell=1)$ & $\tilde{\omega}_2 (\ell=2)$ \\
\hline
0 & 3.0723 & 3.1300\\
0.2 & 2.5872 - 4.4684$\times 10^{-2}$ i & 2.9154 - 5.8491$\times 10^{-5}$ i\\
0.4 & 2.2876 - 0.3951 i & 2.8200 - 2.1218$\times 10^{-2}$ i\\
0.6 & 2.1772 - 0.7998 i & 2.7286 - 0.1226 i\\
0.8 & 2.1392 - 1.1934 i & 2.6826 - 0.2483 i\\
1 & 2.1292 - 1.5823 i  & 2.6573 - 0.3742 i \\
\end{tabular}
\end{ruledtabular}
\end{table}
%%%%%%%%%%%%%%%%%%%%%%%%%%%%%%%%%%%%%%%%%%%%%

\begin{figure}
\begin{center}
\begin{tabular}{c}
\hspace{-4mm}\includegraphics[clip=true,width=0.386\textwidth]{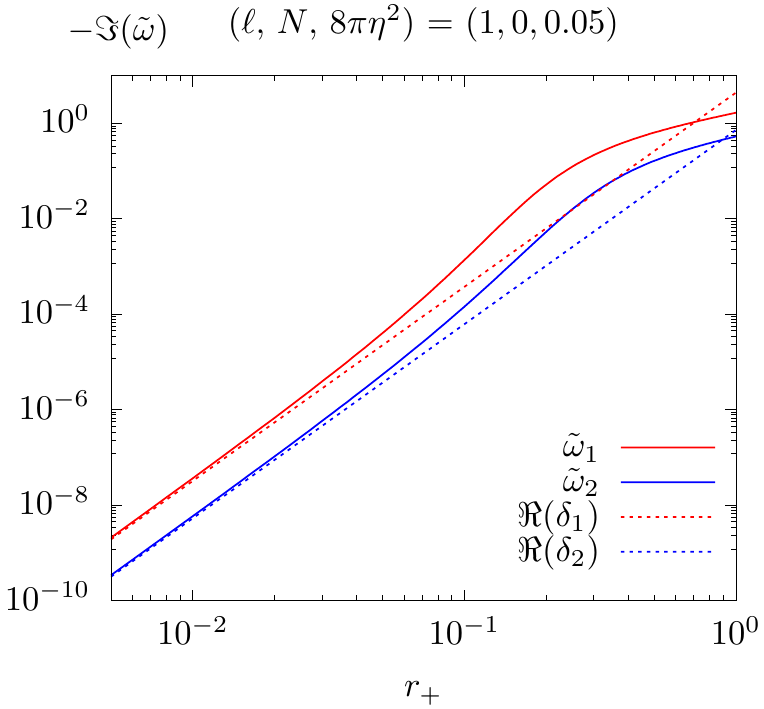}
\end{tabular}
\end{center}
\caption{\label{Fig_comp} Comparison of the imaginary part of quasinormal frequencies for the two sets of fundamental modes with $\ell=1$ and $8\pi\eta^2=0.05$, between the analytic matching approximation (dashed lines) and the numerical data (solid lines). Note that we use double logarithmic coordinates in this figure.}
\end{figure}

In Fig.~\ref{Fig_comp}, we compare the analytic calculations with numeric data, by taking the angular momentum quantum number $\ell=1$, the overtone number $N=0$ and the monopole parameter $8\pi\eta^2=0.05$, and find a good agreement for small black holes.
%%%%%%%%%%%%%%%%%%%%
%%%%%%%%%%%%%%%%%%%%

A few numeric data are tabulated in Table.~\ref{EMmonopole}. As one may observe, by taking $8\pi\eta^2=0.1$ and $N=0$, the real part of the Maxwell QNMs decreases while the magnitude of the imaginary part increases as the black hole size $r_+$ increases, similarly to the Schwarzschild-AdS case. In particular, the isospectrality of the modes for $\ell=1$ with the first boundary condition and $\ell=2$ with the second boundary condition is broken, due to the presence of the global monopole.
  
\begin{figure}
\begin{center}
\begin{tabular}{c}
\hspace{-4mm}\includegraphics[clip=true,width=0.356\textwidth]{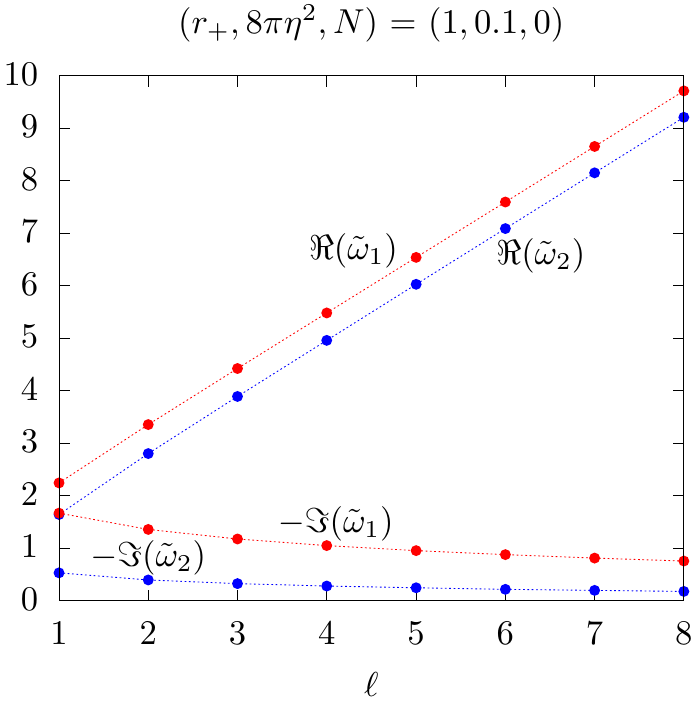}
\end{tabular}
\end{center}
\caption{\label{Fig_elleffects} The impact of the angular momentum quantum number $\ell$ on the real and imaginary parts of quasinormal modes for the first (red) and second (blue) boundary conditions.}
\end{figure}

The effect of the angular momentum quantum number $\ell$ on the Maxwell quasinormal spectrum is presented In Fig.~\ref{Fig_elleffects}, for a black hole with size $r_+=1$, the global monopole $8\pi\eta^2=0.1$ and with the overtone number $N=0$. We observe, similarly to the Schwarzschild-AdS case (i.e. $8\pi\eta^2=0$) reported in~\cite{Wang:2015goa}, that for both boundary conditions the real part of the Maxwell quasinormal frequencies increases while the magnitude of imaginary part decreases, as $\ell$ increases. 
%%%%%%%%%%%%%%%%%%%%
%%%%%%%%%%%%%%%%%%%%

\begin{figure*}
\begin{center}
\begin{tabular}{c}
\hspace{-4mm}\includegraphics[clip=true,width=0.363\textwidth]{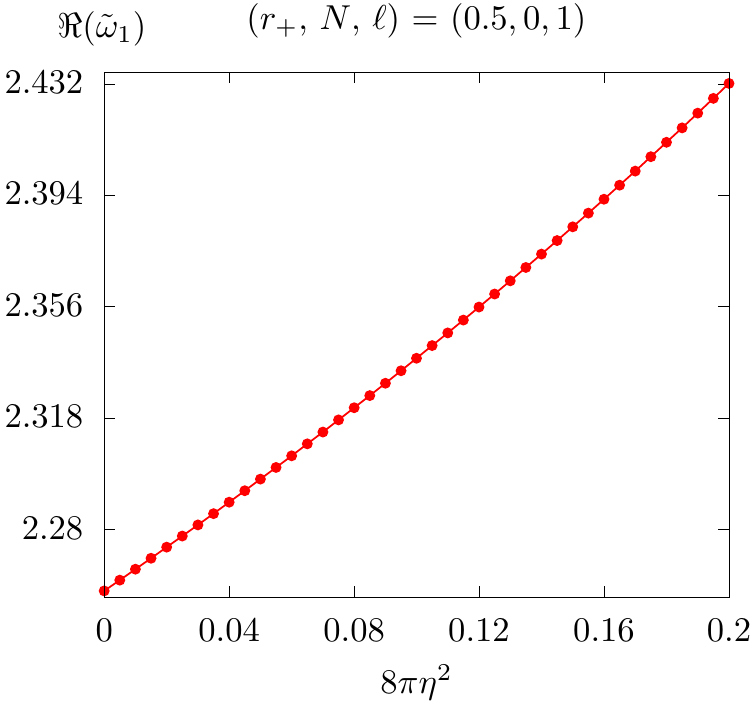}\hspace{12mm}\includegraphics[clip=true,width=0.363\textwidth]{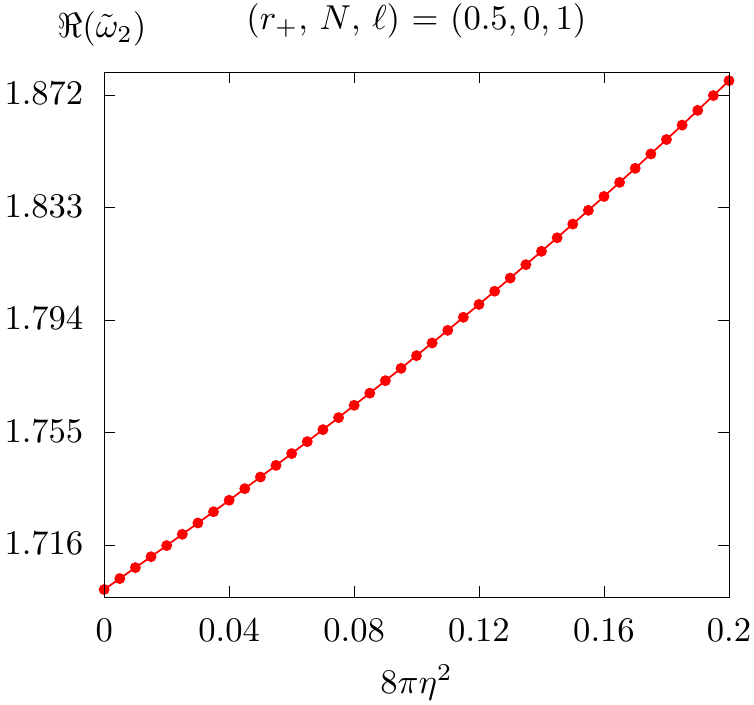}
\vspace{6mm}
\\
\hspace{-2mm}\includegraphics[clip=true,width=0.363\textwidth]{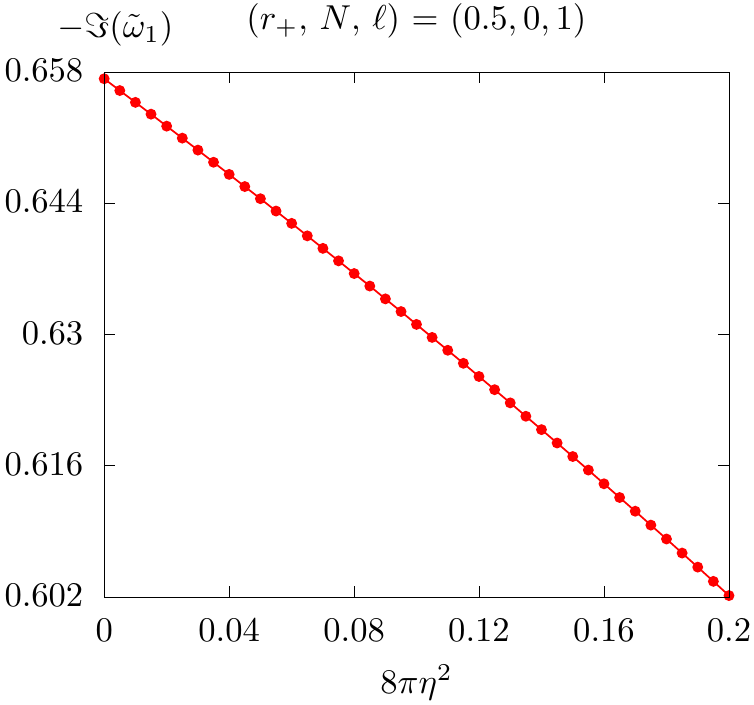}\hspace{12.5mm}\includegraphics[clip=true,width=0.363\textwidth]{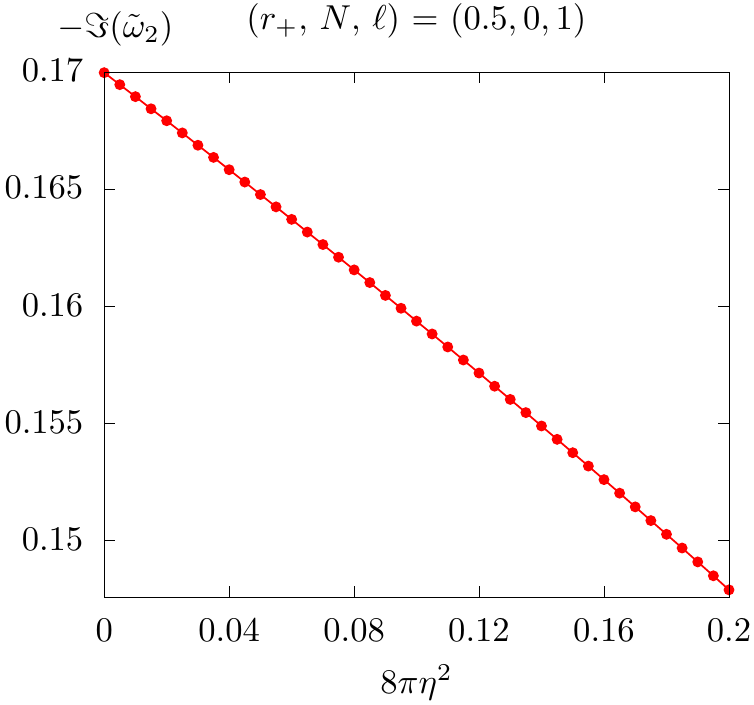}
\end{tabular}
\end{center}
\caption{\label{Fig_monopole} The monopole effects on the real (top) and imaginary (bottom) part of quasinormal spectrum with the first (left) and second (right) boundary conditions, by taking $r_+=0.5$, $\ell=1$ and $N=0$ as an illustrative example. The similar behaviors are also observed for other values of $r_+$, $\ell$ and $N$.}
\end{figure*}

\begin{figure}
\begin{center}
\begin{tabular}{c}
\hspace{-4mm}\includegraphics[clip=true,width=0.36\textwidth]{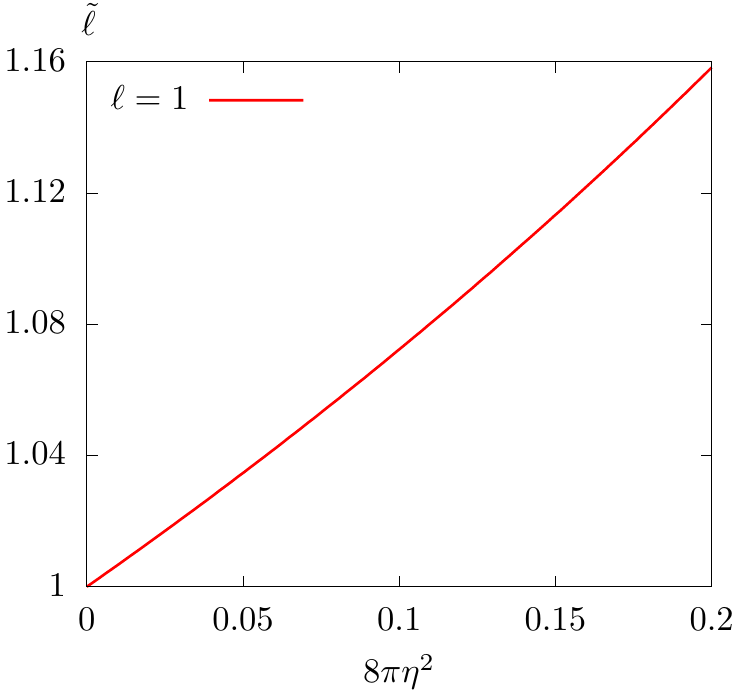}
\end{tabular}
\end{center}
\caption{\label{Fig_ellt} Variation of $\tilde{\ell}$ with respect to the monopole parameter $8\pi\eta^2$, by taking $\ell=1$ as an example. It shows clearly that $\tilde{\ell}$ increases as the monopole parameter  $8\pi\eta^2$ increases.}
\end{figure}

As the main goal of this paper, we explore the impact of the global monopole on the Maxwell spectrum in Fig.~\ref{Fig_monopole}. As an illustrative example, here we take $r_+=0.5$, $\ell=1$, $N=0$, and we observe that, for both boundary conditions, the real (the magnitude of imaginary) part of the Maxwell QNMs increases (decreases) as the global monopole $8\pi\eta^2$ increases. As we have checked for various values of $r_+$, $\ell$ and $N$, the above mentioned behaviors are held. This may be understood as follows. From Eq.~\eqref{RWZeq3}, it shows clearly that the monopole parameter only appears in $\tilde{\ell}$ and $\tilde{\ell}$ plays the same role as $\ell$. From Fig.~\ref{Fig_ellt}, we observe $\tilde{\ell}$, by fixing $\ell$, increases as the monopole parameter $8\pi\eta^2$ increases, indicating the global monopole produces the repulsive force. This implies that, for larger monopole parameter, the perturbation fields (the Maxwell fields here) live longer around black holes, i.e. decay slower, exactly as shown in Fig.~\ref{Fig_monopole}. 

%%%%%%%%%%%%%%%%%%%%
%%%%%%%%%%%%%%%%%%%%%%%%%%%%%%%%%%%%%%%%%%
%%%%%%%%%%%%%%%%%%%%%%%%%%%%%%%%%%%%%%%%%%

\begin{figure*}
\begin{center}
\begin{tabular}{c}
\hspace{-4mm}\includegraphics[clip=true,width=0.3\textwidth]{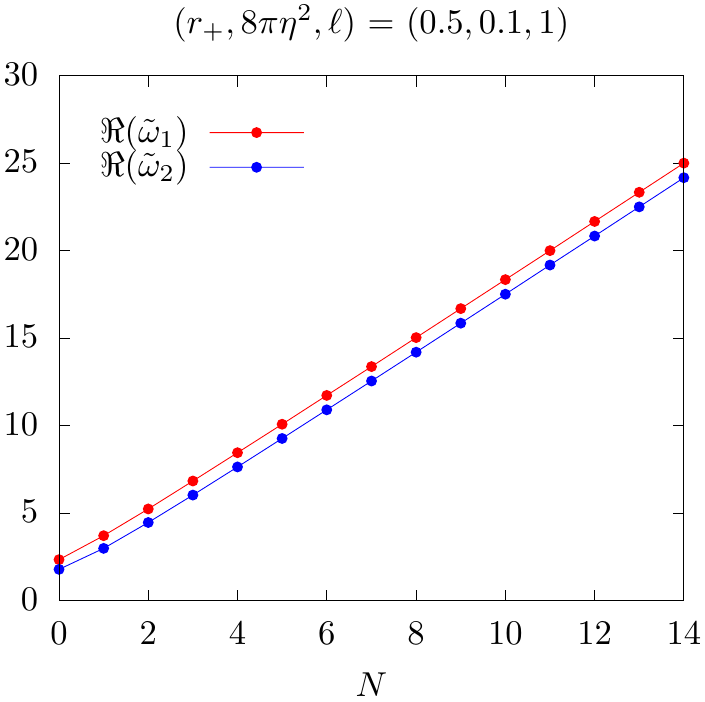}\;\;\;\hspace{2mm}\includegraphics[clip=true,width=0.3\textwidth]{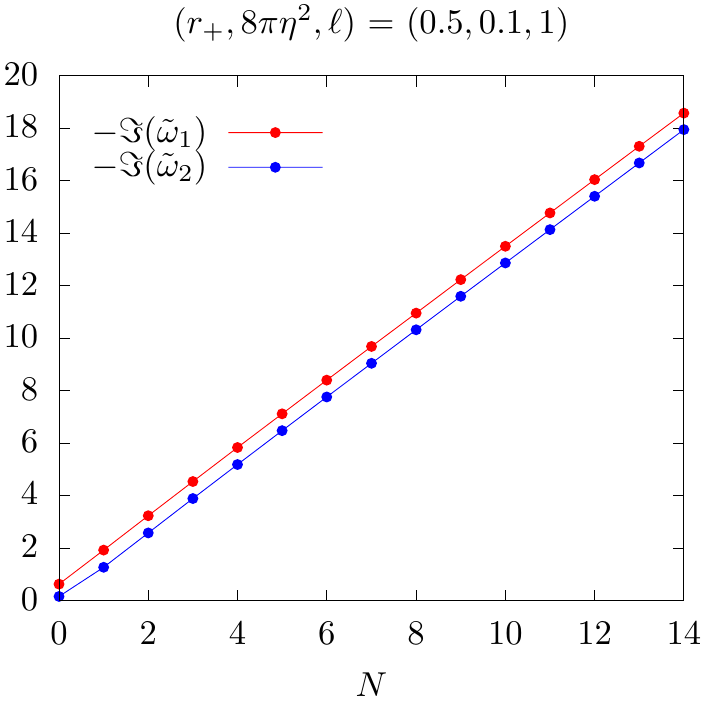}\;\;\;\hspace{2mm}
\includegraphics[clip=true,width=0.3\textwidth]{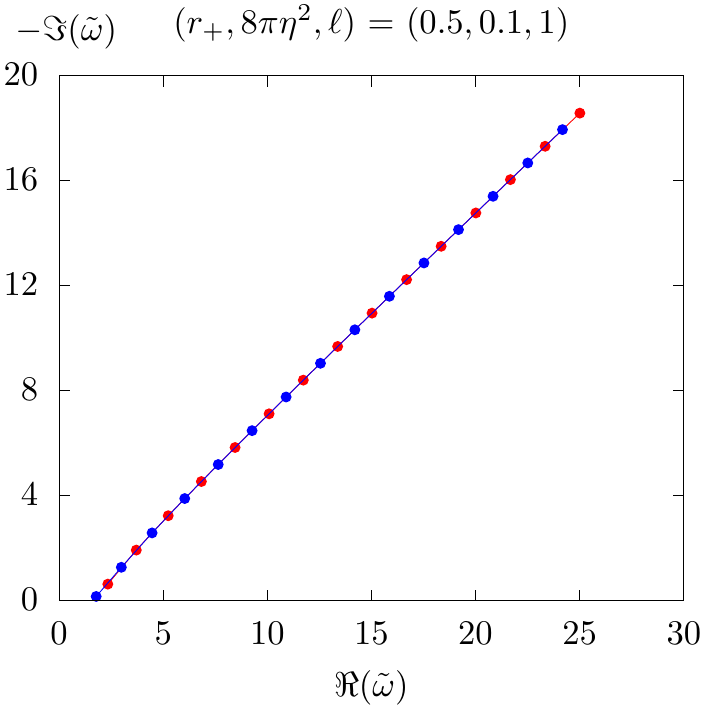}
\end{tabular}
\end{center}
\caption{\label{Fig_Neffects} The impact of the overtone number $N$ on the real (left) and imaginary (middle) parts of quasinormal modes for the first (red) and second (blue) boundary conditions, with fixed $r_+=0.5$, $8\pi\eta^2=0.1$ and $\ell=1$. We also present the imaginary part in terms of the real part of QNMs in the right panel.}
\end{figure*}

We have also studied the dependence of the Maxwell quasinormal frequencies on the overtone number $N$ in Fig.~\ref{Fig_Neffects}. For this case, we take $r_+=0.5$ and $\ell=1$. It is shown, from the left and middle panels, that for two boundary conditions, both the real part and the magnitude of imaginary part of the Maxwell frequencies increase as $N$ increases, and the excited modes for both branches are approximately evenly spaced in N. In the right panel, we display the imaginary part in terms of the real part of the Maxwell QNMs. It shows interestingly that two branches of QNMs (for excited states) lie on the same line for different N. This phenomenon has also been observed for the Dirac case~\cite{Wang:2019qja}, and indicates that, although two branches of QNMs are different, they are similar in the sense that the excited modes of one branch may be interpolated from the other branch.
%%%%%%%%%%%%%%%%%%%%
%%%%%%%%%%%%%%%%%%%%
%%%%%%%%%%%%%%%%%
\section{Discussion and Final Remarks}
\label{discussion}
%%%%%%%%%%%%%%%%%
In this paper we have studied the Maxwell quasinormal spectrum on a global monopole Schwarzschild-AdS black hole, by imposing a generic Robin type boundary condition. To this end, we first presented the Maxwell equations both in the Regge-Wheeler-Zerilli and in the Teukolsky formalisms and derived the explicit boundary conditions for the Regge-Wheeler-Zerilli and the Teukolsky variables, based on the vanishing energy flux principle. Then the Maxwell equations were solved in each formalism, both analytically and numerically.

In a pure AdS space with a global monopole, we have solved the Maxwell equations analytically in the aforementioned two formalisms. We found that two boundary conditions in each formalism lead to two \textit{different} normal modes, due to the presence of the global monopole. This is very different with the Schwarzschild-AdS case where normal modes obtained from two boundary conditions are the same, up to one mode. In the small black hole and low frequency approximations, we also solved the Maxwell equations in the Teukolsky formalism by using an analytic matching method and we verified that the analytic calculations coincide with the numeric data well. 

We then varied the black hole size $r_+$, the angular momentum quantum number $\ell$, and the overtone number $N$, in the presence of a global monopole; and analyzed their effects on the two sets of the Maxwell quasinormal spectrum in the numeric calculations. We observed that, the impact of $r_+$, $\ell$ and $N$ on the Maxwell QNMs are very similar to the Schwarzschild-AdS case. In particular, we explored the monopole effects on the Maxwell spectrum, and we found that for both boundary conditions, the real part of the Maxwell spectrum increases while the magnitude of imaginary part decreases as the monopole parameter $8\pi\eta^2$ increases. These trends are direct consequences of the fact that the global monopole produces the repulsive force.

Finally, we would like to stress that the above mentioned QNMs behaviors were obtained in the unit of $\tilde{L}$. One may alternatively use the unit of $L$, and as we have checked for this case the monopole effects on the Maxwell spectrum are more involved. In the former choice, the Maxwell equations on Schwarzschild-AdS black holes with a global monopole may be reformulated to the Maxwell equations without a global monopole but with the modified angular momentum quantum number $\tilde{\ell}$, so that the repulsive nature of the global monopole becomes more clear.

\bigskip

%%%%%%%%%%%%%%%%%%%%%%%%%%%%%%%%%%%%%%%%%%%%%%%
\noindent{\bf{\em Acknowledgements.}}
%%%%%%%%%%%%%%%%%%%%%%%%%%%%%%%%%%%%%%%%%%%%%%%
This work is supported by the National Natural Science Foundation of China under Grant Nos. 11705054, 11881240252, 11775076, 11875025, 12035005, and by the Hunan Provincial Natural Science Foundation of China under Grant Nos. 2018JJ3326 and 2016JJ1012.

\bibliographystyle{h-physrev4}
\bibliography{MaxwellGlobalAdS_arXiv}

%\newpage

\end{document}